\documentclass[%
 aip,
 amsmath,amssymb,
 reprint,%
]{revtex4-1}

\usepackage{graphicx}% Include figure files
\usepackage{dcolumn}% Align table columns on decimal point
\usepackage{bm}% bold math
\usepackage[colorlinks,linkcolor=blue,citecolor=blue,urlcolor=blue]{hyperref}
\usepackage[utf8]{inputenc}
\usepackage[T1]{fontenc}
\usepackage{mathptmx}
\usepackage{etoolbox}

\makeatletter
\def\@email#1#2{%
 \endgroup
 \patchcmd{\titleblock@produce}
  {\frontmatter@RRAPformat}
  {\frontmatter@RRAPformat{\produce@RRAP{*#1\href{mailto:#2}{#2}}}\frontmatter@RRAPformat}
  {}{}
}%
\makeatother
\begin{document}

\preprint{AIP/123-QED}

\title{Time- and Polarization-Resolved Extreme Ultraviolet Momentum Microscopy}

\author{Sotirios Fragkos}
\affiliation{Universit\'e de Bordeaux - CNRS - CEA, CELIA, UMR5107, F33405 Talence, France}

\author{Quentin Courtade}
\affiliation{Universit\'e de Bordeaux - CNRS - CEA, CELIA, UMR5107, F33405 Talence, France}

\author{Olena Tkach}
\affiliation{Johannes Gutenberg-Universität, Institut für Physik, D-55099 Mainz, Germany}
\affiliation{Sumy State University, Rymskogo-Korsakova 2, 40007 Sumy, Ukraine}

\author{Jérôme Gaudin}
\affiliation{Universit\'e de Bordeaux - CNRS - CEA, CELIA, UMR5107, F33405 Talence, France}

\author{Dominique Descamps}
\affiliation{Universit\'e de Bordeaux - CNRS - CEA, CELIA, UMR5107, F33405 Talence, France}

\author{Guillaume Barrette}
\affiliation{Universit\'e de Bordeaux - CNRS - CEA, CELIA, UMR5107, F33405 Talence, France}

\author{Stéphane Petit}
\affiliation{Universit\'e de Bordeaux - CNRS - CEA, CELIA, UMR5107, F33405 Talence, France}

\author{Gerd Schönhense}
\affiliation{Johannes Gutenberg-Universität, Institut für Physik, D-55099 Mainz, Germany}

\author{Yann Mairesse}
\affiliation{Universit\'e de Bordeaux - CNRS - CEA, CELIA, UMR5107, F33405 Talence, France}

\author{Samuel Beaulieu}
\email{samuel.beaulieu@u-bordeaux.fr}
\affiliation{Universit\'e de Bordeaux - CNRS - CEA, CELIA, UMR5107, F33405 Talence, France}
\date{\today}

\begin{abstract}
We report the development of an instrument combining an ultrafast, high-repetition-rate, polarization-tunable monochromatic extreme ultraviolet (XUV, 21.6~eV) beamline and a next-generation momentum microscope endstation. This setup enables time- and angle-resolved photoemission spectroscopy of quantum materials, offering multimodal photoemission dichroism capabilities. The momentum microscope simultaneously detects the full surface Brillouin zone over an extended binding energy range. It is equipped with advanced electron optics, including a new type of front lens that supports multiple operational modes. Enhanced spatial resolution is achieved by combining the small XUV beam footprint (33~$\mu$m × 45~$\mu$m) with the selection of small regions of interest using apertures positioned in the Gaussian plane of the momentum microscope. This instrument achieves an energy resolution of 44~meV and a temporal resolution of 144~fs. We demonstrate the capability to perform linear, Fourier, and circular dichroism in photoelectron angular distributions from photoexcited 2D materials. This functionality paves the way for time-, energy-, and momentum-resolved investigations of orbital and quantum geometrical properties underlying electronic structures of quantum materials driven out of equilibrium.
\end{abstract}

\maketitle

\section{\label{sec:intro}Introduction}

Angle-resolved photoemission spectroscopy (ARPES) is widely regarded as one of the most advanced techniques for investigating the electronic properties of quantum materials~\cite{Sobota21, Lv19}. By leveraging energy and momentum conservation laws inherent to the photoelectric effect, ARPES enables the determination of electrons' binding energies and in-plane momenta, facilitating detailed mapping of the electronic band structure of crystalline solids. In recent years, various flavors of ARPES have emerged to address the specific demands of rapidly expanding areas in quantum materials research. These advancements include improvements in spatial resolution (micro- and nano-ARPES)~\cite{Bostwick12, Avila14, Rotenberg14, Koch18}, the exploitation of transition dipole matrix element effects through polarization control~\cite{Sterzi18, Volckaert19, Rostami19, Min19, Moser22, Cho18, Cho21, Schuler20-1, Beaulieu20, Beaulieu21-3, Schuler22}, imaging spin degrees of freedom (spin-ARPES)~\cite{Meier09, Jozwiak10, Souma10, Berntsen10, Okuda11, Kolbe11}, accessing three-dimensional band structures through photon energy tunability~\cite{Saitoh2000-bk, Reininger2012-jh, Tamura2012-te, Strocov2014-sg}, and probing out-of-equilibrium properties of solids using pump-probe schemes (time-resolved ARPES, trARPES)~\cite{Boschini2024-jj, Mathias2007-se, Carpene2009-zw, Dakovski2010-sg, Berntsen2011-lf, Smallwood2012-dr, Frietsch13, Ishida2014-ln, Ishida2016-aw, Jiang2014-jj, Rohde16, Keunecke2020-ym, Huang2022-xv}. Integrating all these ARPES modalities into a single instrument would provide unprecedented versatility for exploring various facets of the electronic structure of quantum materials. While such an instrument has yet to be fully realized, a significant trend in the field is the coupling of different ARPES variants to uncover previously inaccessible information. For instance, spin- and time-resolved ARPES (STARPES) enables the study of ultrafast spin-polarization dynamics~\cite{Schonhense2015, Gotlieb13, Plotzing2016-ga, Fanciulli20, Nie19}. Similarly, other combinations, such as micro- and time-resolved ARPES~\cite{Dufresne24} or photon energy tunable trARPES~\cite{Sie19, Guo22, Pan2024-gv}, have opened new avenues for exploring the electronic properties of quantum materials.

Time-resolved ARPES (trARPES) has rapidly evolved into a powerful technique for investigating nonequilibrium properties of quantum materials, driven by a series of technological innovations~\cite{Na2023-eh, Boschini2024-jj}. Key advancements include the optimization of femtosecond laser systems, harmonic generation in nonlinear crystals, high-harmonic generation (HHG) in noble gases, and significant improvements in photoelectron detection technology. Although extreme-ultraviolet (XUV) pulses generated via HHG have been employed in photoemission spectroscopy since the 1990s~\cite{Haight1994-sy}, for decades, most trARPES systems utilized $\sim$6~eV UV probe pulses, typically produced either by quadrupling the output of femtosecond Ti:Sa lasers or through sum-frequency generation of Yb-based laser systems~\cite{Gauthier2020-hz, Carpene2009-zw, Faure2012-ej, Smallwood2012-dr, Boschini2014-hw}. The straightforward design and accessibility of these low-photon-energy systems enabled early studies of ultrafast electron dynamics in quantum materials such as topological insulators~\cite{Wang2013-dk, Mahmood2016-jv, Zhu2015-ks, Sanchez-Barriga2016-vr, Jozwiak2016-vm}. Additionally, light polarization in this UV spectral range can be easily manipulated using standard half- and quarter-wave plates, allowing for the integration of time-resolved UV ARPES with dichroic ARPES~\cite{Niesner2012-th, Mauchain2013-pe, Zhang2021-ib, Hedayat2021-dw}. This combination has been proven valuable for probing symmetry-related properties in the electronic structures of photoexcited solids. Despite these advances, UV probe photons generated via nonlinear crystals have an inherent limitation: their restricted access to electrons with small in-plane momenta. This constraint confines measurements primarily to electron dynamics near the Brillouin zone center.

\begin{figure*}[t]
\begin{center}
\includegraphics[width=17cm]{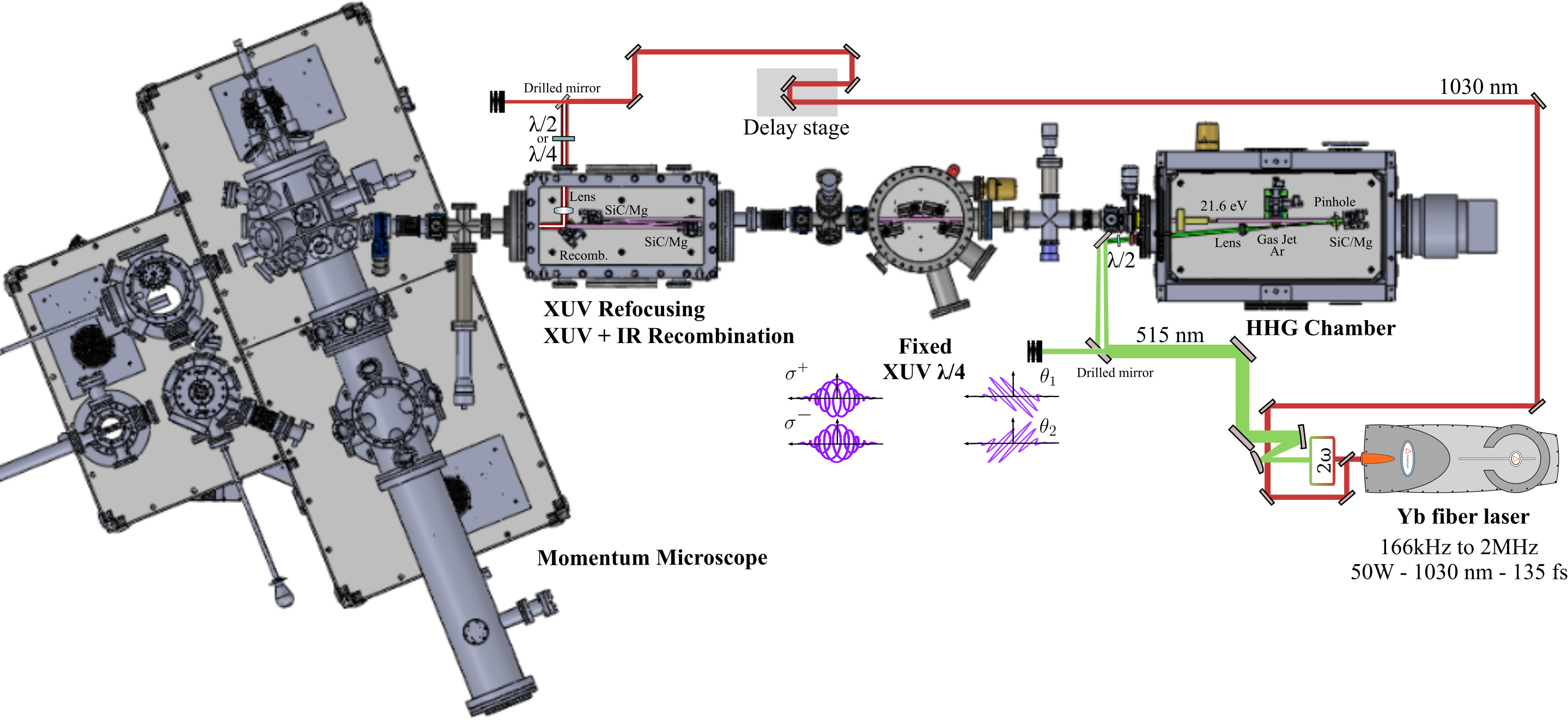}
\caption{\textbf{Schematic layout of the time- and polarization-resolved extreme ultraviolet momentum microscopy apparatus.} A scheme of the full beamline including the amplified Yb fiber laser, the frequency doubling module, the high-order harmonic generation chamber, the fixed all-reflective XUV quarter-wave plate, the XUV/IR recombination and refocusing chamber, as well as the momentum microscope endstation. In both pump and probe arms, beams are actively stabilized just before entering vacuum chambers using two-quadrant position sensors feeding two mirrors controlled with piezoelectric adjusters, ensuring long-term alignment stability.}
\label{fig:schematic}
\end{center}
\end{figure*}

To achieve full coverage of the Brillouin zone across all materials, photon energies exceeding $\sim$20~eV are required, i.e. within the XUV spectral range~\cite{Mathias2007-se, Dakovski2010-sg, Heyl2012-fr, Frietsch13, Eich14, Puppin2019-ra, Plotzing2016-ga, Rohde16, Buss19, Heber2022-zn, Zhong2024-ki}. Femtosecond XUV pulses, generated via HHG in rare gases~\cite{McPherson87, Ferray88}, enable the study of ultrafast electron dynamics across the entire momentum space of quantum materials. The repetition rate of HHG sources was long limited to a few kHz by the femtosecond Ti:Sa laser technology. The development of high-power femtosecond Yb-based laser systems in the 2000s has changed the game, enabling HHG at MHz repetition rate~\cite{Boullet09}. The reliability of industrial-grade Yb lasers has made such XUV sources integrated with ARPES detectors increasingly accessible~\cite{Na2023-eh, Chiang2012-bi, Corder18, Mills2019-rx, Sie19, Puppin2019-ra, Lee20, Peli2020-ns, Guo22}. These technological breakthroughs now enable high-repetition-rate XUV trARPES to be combined with other ARPES modalities, such as spin-resolved or dichroic ARPES.

Controlling light-matter interaction through the modulation of the light polarization state allows the extraction of dichroic observables, such as linear dichroism in the photoelectron angular distributions (LDAD)~\cite{Sterzi18, Volckaert19, Rostami19, Min19, Unzelmann20, Moser22, Schusser2022} and circular dichroism in the photoelectron angular distributions (CDAD)~\cite{Wang2011-aj, Liu2011-tr, Lin2018-hv, Fedchenko2019-da, Cho18, Cho21, Unzelmann2021, Schuler20-1, Schusser2022, Brinkman24, figgemeier2024}. Recently, new dichroic observables, including intrinsic linear~\cite{Beaulieu21-3}, time-reversal~\cite{beaulieu20-2, Schusser2024}, and Fourier dichroism~\cite{Schuler22} in the photoelectron angular distributions (\textit{i}LDAD, TRDAD, and FDAD, respectively), have been introduced. These dichroic observables are known to be sensitive to the orbital texture and quantum geometric properties of the wavefunction underlying the electronic structure of solids, even though experimental geometry, final-state effects, and scattering can complicate the interpretation of the origin of these signals~\cite{Schonhense90, moser23, boban2024}. Combining these dichroic observables with XUV trARPES would enable real-time tracking of the ultrafast evolution of the local Berry curvature, orbital pseudospin, and topology of the electronic band structure undergoing dynamics in photoexcited solids~\cite{Schuler20-5, Schuler2022}. However, a fully polarization-tunable XUV trARPES setup for time-resolved photoemission dichroism in solids driven out of equilibrium has not yet been reported. 

In this work, we present the coupling of a high-repetition-rate polarization-tunable monochromatic XUV source with a momentum microscopy apparatus. Detailed characterization of the instrument is provided, demonstrating its capabilities for time- and angle-resolved photoemission spectroscopy of quantum materials, as well as its potential for multimodal photoemission dichroism studies.

\section{\label{sec:Beamline}XUV Beamline Overview}

The XUV beamline is composed of five main building blocks (see Fig.~\ref{fig:schematic}): \textit{i}) the amplified Yb fiber laser, whose beam is split in two arms -- pump and probe, \textit{ii}) the frequency doubling and HHG chamber for XUV pulse generation, \textit{iii}) the all-reflective XUV quarter-wave plate, \textit{iv}) the XUV refocusing and \textit{v}) the polarization manipulation, delay adjustment, focusing, and recombination of pump pulses. The development of building blocks \textit{i})--\textit{iii}), along with a detailed discussion of the optimization and scaling of the XUV flux, spectral selection, and XUV polarization state characterization, has been described in a previous publication~\cite{Comby22}. 

This section of the manuscript will briefly describe all relevant aspects of the XUV beamline, emphasizing the design principles and characteristics required for its coupling with the momentum microscope. More details about the characterization of the beamline's spatial, temporal, and spectral resolution will be provided in the following sections. 

The XUV beamline is driven by the BlastBeat femtosecond laser system (Tangerine Short Pulse, Amplitude Group) at the Centre Lasers Intenses et Applications (CELIA) in Bordeaux, France. This laser system consists of two Yb-doped fiber amplifiers seeded by the same oscillator, delivering two synchronized 50~W beams of 135~fs pulses centered at 1030~nm with a spectral bandwidth of 18.5~nm (FWHM). The present experimental setup uses one of these two amplifiers. The repetition rate is continuously tunable between 166~kHz and 2~MHz. The beamline was designed for a laser operation between 166~kHz and 250~kHz. All the data presented in this manuscript were obtained at a repetition rate of 166~kHz.

The laser beam pointing is actively stabilized (MRC Systems GmbH) before the beam is split in two by a polarizing beamsplitter. The reflected beam ($\sim$28~W) is used for the pump arm. We typically use only tens to hundreds of mW from this high-power IR beam to pump the sample. The transmitted beam ($\sim$22~W), dedicated to the probe arm, is frequency doubled using a 1~mm-thick BBO crystal (Castech) to generate up to 7~W at 515~nm. The BBO crystal is housed in an oven (Fastlite) maintained at approximately 120$^\circ$C to prevent long-term degradation (e.g., fogging) caused by its hygroscopic nature. The 515~nm driving beam is expanded using a $-75~\text{mm}/+300~\text{mm}$ ($\times 4$) telescope before being reflected off a drilled mirror (hole diameter $\phi = 4~\text{mm}$). Subsequently, the beam is reduced using a $+125~\text{mm}/-75~\text{mm}$ ($\times 0.6$) telescope. These two telescopes, positioned before and after the drilled mirror, enable precise adjustment of the sizes of the inner and outer regions of the annular beam. This annular beam, which retains 80$\%$ of the initial laser power, facilitates the spatial separation of the intense visible driver from the XUV beamlet.

Before entering the HHG vacuum chamber, the beam position and pointing are actively stabilized (Polaris and K-Cube, Thorlabs) to correct for drifts, and a motorized (motorized rotation stage, Newport) zero-order half-wave plate (B. Halle) is used to control the linear polarization axis angle of the driving 515~nm pulses. The pulses are focused with an $f = 79$~mm aspheric lens into a thin and dense argon gas jet (glass tube with 50~$\mu$m diameter and conical tip, Hilgenberg). A movable gas catcher positioned above the gas jet directly extracts portions of the argon gas, reducing XUV reabsorption by the residual gas. A pinhole located $\sim$190~mm after the gas jet blocks the annular 515~nm beam and transmits the XUV beamlet propagating within the hole of the annular beam, while an $f = 200$~mm SiC/Mg multilayer mirror collimates the XUV beam (maximum reflectivity 40\% at 21.6~eV, partially suppressing adjacent harmonics, NTT-AT). The XUV mirror mount is motorized (Newfocus) to adjust the beam direction. Spectral selection is further refined using a Sn foil (200~nm thick, Luxel) with 22\% theoretical transmission at 21.6~eV. 

The XUV beam can then pass through an XUV quarter-wave plate (QWP) made of four SiC mirrors (NTT-AT) under $78^\circ$ incidence, designed for $\lambda/4$ phase retardance at 21.6~eV, with 41\% transmission in s-polarization and 21\% in p-polarization. This enables the generation of quasi-circular XUV pulses (ellipticity up to 90\%~\cite{Comby22}) for circular dichroism experiments. The XUV-QWP is installed on a motorized translation stage (Newport), allowing it to be easily inserted or removed from the XUV beam path. This design permits the utilization of the total photon flux for experiments using a linearly polarized XUV probe.

\begin{figure}[t]
\begin{center}
\includegraphics[width=8.6cm]{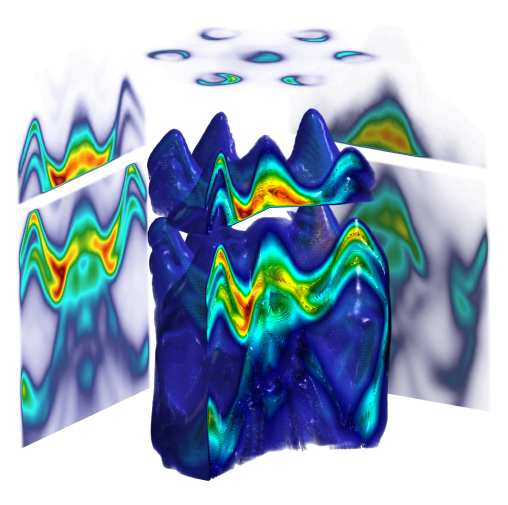}
\caption{\textbf{Exemplary data generated with our momentum microscope apparatus.} Three-dimensional photoemission intensity $I(E_\mathrm{B}, k_x, k_y)$ from 2H-WSe$_2$, acquired using circularly-polarized driving IR pump pulse (1.2~eV, 135~fs, $\sim$5.7~mJ/cm$^2$) and while continuously rotating the linear polarization axis angle of the XUV probe beam (21.6~eV) at the pump-probe temporal overlap. The photoemission intensity is integrated for all XUV polarization axis angles. The light scattering plane was aligned with the crystal mirror plane, i.e. along the $\Gamma$-M direction. The signal above the valence band maximum has been multiplied by 15 to enhance the visibility of the pump-induced Volkov states. The cut on the left- (right-) hand side is along K-$\Gamma$-K$^{\prime}$ (M-$\Gamma$-M$^{\prime}$) high symmetry direction, while the constant energy contour above the 3D photoemission intensity is taken at the energy of the first-order replica of the first valence band (Volkov states).}
\label{fig:3D}
\end{center}
\end{figure}

The polarization-tunable XUV beam then propagates towards the XUV refocusing and recombination chamber. It first impinges on a $f = \text{-250~mm}$ concave mirror near normal incidence and then onto a $f = \text{+350~mm}$ convex mirror, also near normal incidence, which refocuses the beam with an effective focal length of $\sim$1000~mm, which coincides with the sample position in the analysis chamber of the momentum microscope. The two concave/convex mirrors also have a SiC/Mg multilayer coating (NTT-AT) with a maximum reflectivity of 40\% at 21.6~eV, thus enhancing the spectral contrast with adjacent harmonics. These two XUV mirror mounts are also motorized (Newfocus) to finely adjust the beam alignment. This focusing telescope ensures a sub-50~$\mu$m FWHM XUV beam footprint on the sample, which is smaller than in most time-resolved XUV ARPES setups previously reported in the literature. More details about in-situ characterization of beam's footprints using the PhotoElectron Emission Microscopy (PEEM) mode of our momentum microscope will be given in section~\ref{sec:PEEM}. The XUV refocusing chamber also hosts the optics to collinearly recombine the XUV probe beam with the IR pump. 

Concerning the IR pump arm, the fraction of the 1030~nm fundamental laser picked up by the polarizing beamsplitter is directed onto a motorized delay stage (Newport). The beam is reflected off a drilled IR mirror (hole diameter $\phi = 5~\text{mm}$) and its position and pointing are actively stabilized (Polaris and K-Cube, Thorlabs) before entering the refocusing/recombination chamber. The annular IR pump beam goes through a drilled lens ($f = \text{+750~mm}$) before being reflected by a drilled IR mirror (hole diameter $\phi = 3~\text{mm}$) at 45$^{\circ}$, which acts as an XUV/IR collinear recombination optics. Indeed, while the XUV beam propagates in the hole of the drilled mirror, the converging IR beam is reflected off the surface of the mirror. The two beams are thus collinearly recombined and can be spatially and temporally overlapped onto the sample surface in the analysis chamber of the momentum microscope. The polarization state of the IR pump is controlled by a motorized quarter or half waveplate. Polarimetry measurements were conducted after the 45$^\circ$ recombination mirror to estimate the influence of the phase shift between s- and p- polarization components on the polarization state of the reflected light and showed a negligible effect.

\section{\label{sec:MM}Momentum Microscope}

The hemispherical analyzer is the most widely used detector in time-resolved ARPES ~\cite{maklar20, Hufner2010-eb}. In this detection scheme, photoelectrons first pass through an electrostatic lens system and are then guided into hemispherical deflector electrodes, which act as a dispersive bandpass energy filter. The electrons are subsequently projected onto a two-dimensional (2D) multi-channel plate (MCP) detector, capturing a narrow 2D slice ($E_{\text{B}}$, $k_x$) of the 3D photoelectron energy-momentum distribution. Recovering the 3D photoelectron distribution ($E_{\text{B}}$, $k_x$, $k_y$) thus requires recording a series of 2D slices, by changing $k_y$ via varying the angle of the sample with respect to the detector axis or by sweeping the electron distribution using deflectors. During data collection, rotation of the sample would be adversarial to any dichroism measurements since the modification of the experimental geometry can have strong impacts on the transition dipole matrix elements. 

The developed time-of-flight momentum microscopes (TOF-MM) overcome this limitation~\cite{Medjanik17, maklar20, Schonhense2015, Tusche16, tkach2024multimodelensmomentummicroscopy}. Indeed, the working principle of a momentum microscope (MM) is based on applying a high positive voltage to an electrostatic objective lens near the sample surface, enabling the collection of photoelectrons from the entire 2$\pi$ solid angle in an energy range of several eV. An example of three-dimensional photoemission intensity from 2H-WSe$_2$ is shown in Fig.~\ref{fig:3D}. In addition, the momentum microscope is equipped with advanced electron optics~\cite{tkach2024multimodelensmomentummicroscopy}, allowing it to create images of the photoemitted electron distribution in both real- and momentum-space. Indeed, analogous to optical microscopy, an image of the surface-projected band structure is created in the back focal plane of the objective lens. Adjustable contrast aperture can be inserted in this image plane (termed reciprocal image plane) to select a given momentum area, enabling dark-field PEEM, i.e. momentum-resolved measurements of the real-space electron distribution~\cite{Barrett2012-in}. Similarly, an adjustable field aperture can be inserted in the intermediate real-space image plane to better define the source area, i.e. down to sub-micron diameter~\cite{Fedchenko22}. The electrons then pass through a dispersive field-free TOF drift tube, and subsequently, their 2D momentum and kinetic energy are recorded using an MCP in front of a position-sensitive delay-line detector (DLD). This type of time-of-flight MM thus enables the simultaneous measurement of the full 3D photoelectron distribution across the entire in-plane momentum range and over a broad energy range. In addition, it can easily be switched between PEEM-mode (with dark-field imaging capabilities) and ARPES-mode (with spatially-resolved band mapping capabilities). 

This elegant detection scheme, unfortunately also comes with some drawbacks. First, high extractor fields used in standard TOF-MM often cause problems due to field emission at corners or protrusions of cleaved or heterogeneous samples. The sample morphology and surface quality requirement for momentum microscopy using classical extractor-type front lens optics is thus enhanced compared to standard hemispherical analyzers. Moreover, detrimental space-charge effects, i.e. strong Coulomb interaction within photoemitted electrons leading to distortions of the angular and energy distribution, appear at lower electron density than in standard hemispherical analyzers~\cite{Schönhense18, maklar20}. Indeed, before the dispersive drift tube, fast direct and slow secondary electrons do not spread significantly along their propagation axis, leading to enhanced interaction length and time. Refocusing the photoelectron disk at several focal planes of the lens column further increases these space-charge effects.

The present TOF-MM is a prototype with improved performance for $k$-imaging. Special features include enlarged arrays of contrast apertures and field apertures, as well as improved lens optics. The enlarged array of 9 contrast apertures with increased transverse motion using piezomotors improves the performance for Gaussian imaging in bright-field and dark-field modes. The array of 16 field apertures enables $k$-imaging with regions-of-interest (ROIs) down to the sub-micron range. Defining the ROI by a field aperture avoids artifacts caused by shifts of the photon footprint, e.g. upon beam pointing drifts or changing the photon energy. 

The improvement on the electron optics side is a new type of front lens that allows multiple modes of operation. Cathode lenses of conventional PEEMs and MMs operate in the extractor mode with a strong homogeneous accelerating field (typically 4-8~kV/mm) between the sample and the extractor. Several materials currently under investigation require cleavage before photoemission experiments, which can result in the formation of sharp corners or protruding lamellae, e.g. in the case of layered materials. When working with such samples, high extractor fields often cause problems due to the potential for field emission at corners or protrusions. Further technical details, such as the enhanced k-resolution and increased k-field of view are discussed in theory~\cite{tkach2024multimodelensmomentummicroscopy} and experimentally~\cite{tkach24}. The repeller mode~\cite{Schönhense21} can provide access to regions of the pump fluence that are inaccessible in the standard mode, as shown in an experiment at the free-electron laser FLASH~\cite{Elmers20}. The design principle and new features of multi-mode lens for momentum microscopy have been discussed in detail elsewhere~\cite{tkach2024multimodelensmomentummicroscopy,tkach24}. 

\section{\label{sec:PEEM}Spatial Resolution using PEEM mode}

\subsection{IR pump and XUV probe beam's footprint}

\begin{figure}[t]
\begin{center}
\includegraphics[width=8.6cm]{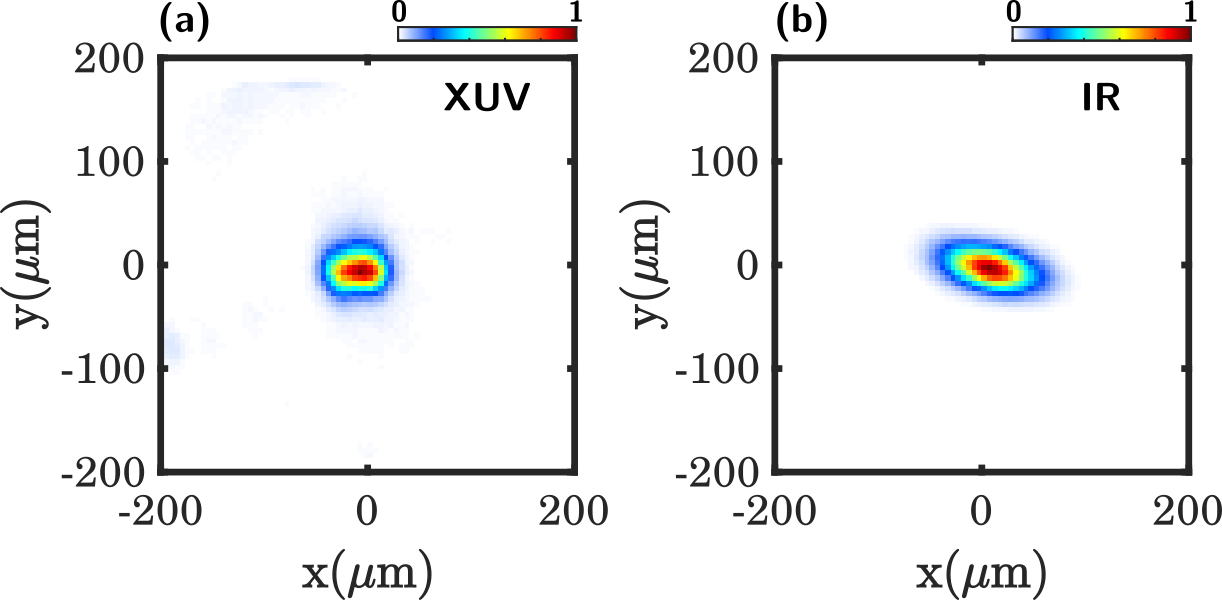}
\caption{\textbf{Characterization of XUV and IR beam footprints on the sample.} \textbf{(a)} PEEM imaging of the XUV (21.6~eV) probe beam footprint. The FWHM footprint is 45~$\mu$m along $x$ and 33~$\mu$m along $y$.  \textbf{(b)} PEEM image of the IR (1.2~eV) pump beam footprint.  The measured FWHM photoemission spot size is 60~$\mu$m along $x$ and 30~$\mu$m along $y$, which is associated with an IR pump beam footprint of 139~$\mu$m along $x$ and 70~$\mu$m along $y$ when taking into account the effective nonlinearity of the multiphoton photoemission process.}
\label{fig:footprint}
\end{center}
\end{figure}

The Photoemission Electron Microscopy (PEEM) mode of our time-of-flight momentum microscope allows us to record real-space images of photoelectrons ejected from the surface. We used the PEEM mode to determine the pump and probe beam footprints onto the surface of a 2H-WSe$_2$ sample. Figure~\ref{fig:footprint} (a) shows real-space image of the XUV probe beam footprint. Since XUV photoemission is a single photon process, the PEEM image directly reflects the XUV beam footprint. Figure~\ref{fig:footprint} (b) shows the real-space image obtained by irradiating the sample with the IR pump beam. Since photoemission using IR (1.2~eV) is a multiphoton process, the PEEM image does not directly reflect the IR beam footprint, as the effective nonlinearity of the process needs to be taken into account. 

We performed Gaussian fits of the XUV and IR beam footprint along the $x$ and $y$ directions. For the XUV probe beam, we find a full-width half maximum (FWHM) footprint of 45~$\mu$m along $x$ and 33~$\mu$m along $y$. The larger footprint along the $x$ direction can be partially explained by the 65$^{\circ}$ angle of incidence of the beam on the crystal surface. For the IR pump beam, we find a FWHM footprint of 60$\mu$m along $x$ and 30$\mu$m along $y$. To measure the effective nonlinearity of the IR photoemission process, we recorded the total photoemission intensity $S_\mathrm{tot}$ as a function of the IR pump power $P_0$. The results show a linear behavior when plotted in log-log scale, indicating that $S_\mathrm{tot}\propto P_0^{\alpha}$, with an effective nonlinearity $\alpha$ = 5.4 extracted by linear fit. The real IR pump beam footprint can thus be retrieved by multiplying the measured footprint using multiphoton PEEM by the square root of the effective nonlinearity. This leads to a real IR pump beam footprint of 139~$\mu$m FHWM along $x$ and 70~$\mu$m along $y$. 

\subsection{Ultimate spatial resolution}

\begin{figure}[t]
\begin{center}
\includegraphics[width=8.6cm]{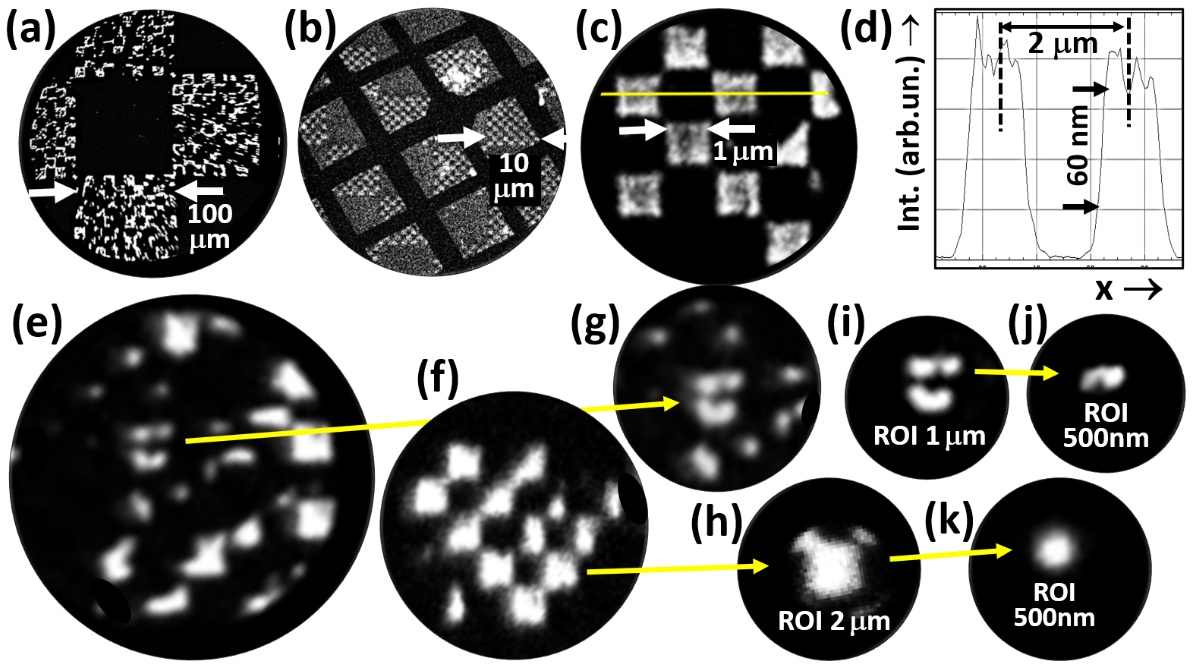}
\caption{\textbf{Selection of small regions-of-interest in PEEM mode.} \textbf{(a-c)} PEEM imaging of a Au on Si test sample at different magnifications. \textbf{(d)} Intensity line scan along the line marked in \textbf{(c)}. \textbf{(e-k)} Region-of-interest (ROI) selections using field apertures in the intermediate image plane. The smallest field apertures select individual 1~$\mu$m squares \textbf{(h,i)} and even sub-micron details within a square \textbf{(j,k)}.}
\label{fig:spatial}
\end{center}
\end{figure}

Next, we performed advanced characterization of the real space imaging mode (PEEM) with a 'Chessy' sample (hierarchical four-fold checkerboard pattern of Au on Si), using excitation by a continuous wave (CW) UV LED (light-emitting diode, 4.73~eV, 262~nm). Figure~\ref{fig:spatial} displays PEEM images at different magnifications, showing the hierarchical checkerboard structure with features on the (a) 100~$\mu$m, (b) 10~$\mu$m, and (c) 1~$\mu$m scale.  All panels of Fig.~\ref{fig:spatial} except (b) were recorded with the UV LED; (b) was recorded with a Hg lamp (photon energy 4.9 eV). Overlaid with the checkerboard pattern, Fig.~\ref{fig:spatial} (b) shows the shadow image of an auxiliary grid used to adjust the first Gaussian image exactly in the plane of the field apertures. This is a necessary condition for the sub-micron region of interest (ROI) selection. The Fig.~\ref{fig:spatial} data were recorded in extractor mode, with the extractor and ring electrode both set to 12 kV.

The Chessy test sample shows a lot of inhomogeneities, which are highlighted by the UV-LED excitation, making it possible to identify individual squares and local defects. The smallest squares have a size of $1 \times 1$~$\mu$m$^2$, see Fig.~\ref{fig:spatial} (c), allowing the lateral resolution to be determined. The intensity profile along the line marked in Fig.~\ref{fig:spatial} (c) is shown in Fig.~\ref{fig:spatial} (d). The edge line scan shows a width of 60~nm. This is an upper limit for the resolution because the finite thickness of the Au patches causes a field deformation that increases the apparent width.

For the selection of small ROIs, one proceeds as follows. First, the focusing is adjusted such that the intermediate Gaussian image is in the plane of the field apertures, see the example in Fig.~\ref{fig:spatial} (b). Here, the magnification ($M$) between the sample coordinates and the intermediate image is approximately $M=10$. The grid in Fig.~\ref{fig:spatial} (b) has a periodicity of 127~$\mu$m, corresponding to 12.7~$\mu$m in sample coordinates; the scale agrees with the image of the Au pattern. Next, the desired field aperture is moved in, selecting the region of interest. The image sequence in Figs.~\ref{fig:spatial} (e-k) shows examples with stepwise reduced ROIs down to sizes of 2~$\mu$m (h), 1~$\mu$m (i), and 500~nm (j,k). At the given magnification of $M=10$, the 500~nm ROI patterns have been obtained with a field aperture of 5~$\mu$m diameter in the intermediate Gaussian image. Switching to the $k$-imaging mode makes momentum patterns of photoelectrons from small ROIs, down to sub-micron sizes, visible. This approach to sub-micron ARPES works independently of the size of the photon footprint.  Using a field aperture with a diameter of 5 $\mu$m, an ROI with a diameter of 900 nm was shown in an earlier experiment~\cite{Fedchenko22}. Further improvement by a factor of two is realistic by optimizing the photon flux in the ROI.

\section{\label{sec:resolutions}Time and energy resolutions}

To characterize the energy resolution, one can use the width of the measured Fermi-energy cutoff. The total resolution $\Delta E_{\text{tot}}$ is determined by the Fermi-Dirac distribution at a given temperature $T$, convolved with the bandwidth of the photoionizing radiation $\Delta E_{h\nu}$ and the resolution of the detector (here $\Delta E_{\text{ToF}}$). The TOF-MM is equipped with a Lakeshore ST-400 Series ultra-high vacuum cryostat, enabling cooling down to $\sim$16~K using liquid helium. The thermal width of the Fermi-Dirac distribution is $\Delta E_{\text{therm}} \approx 4kT$, which is about 100~meV at room temperature. The total resolution $\Delta E_{\text{tot}}$ is given by the squared sum of these contributions:  

\begin{align}
	\label{eq:deltaE}
	\Delta E_{\text{tot}}^2 = \Delta E_{h\nu}^2 + \Delta E_{\text{ToF}}^2 + \Delta E_{\text{therm}}^2.
\end{align}

\begin{figure}[t]
\begin{center}
\includegraphics[width=8.6cm]{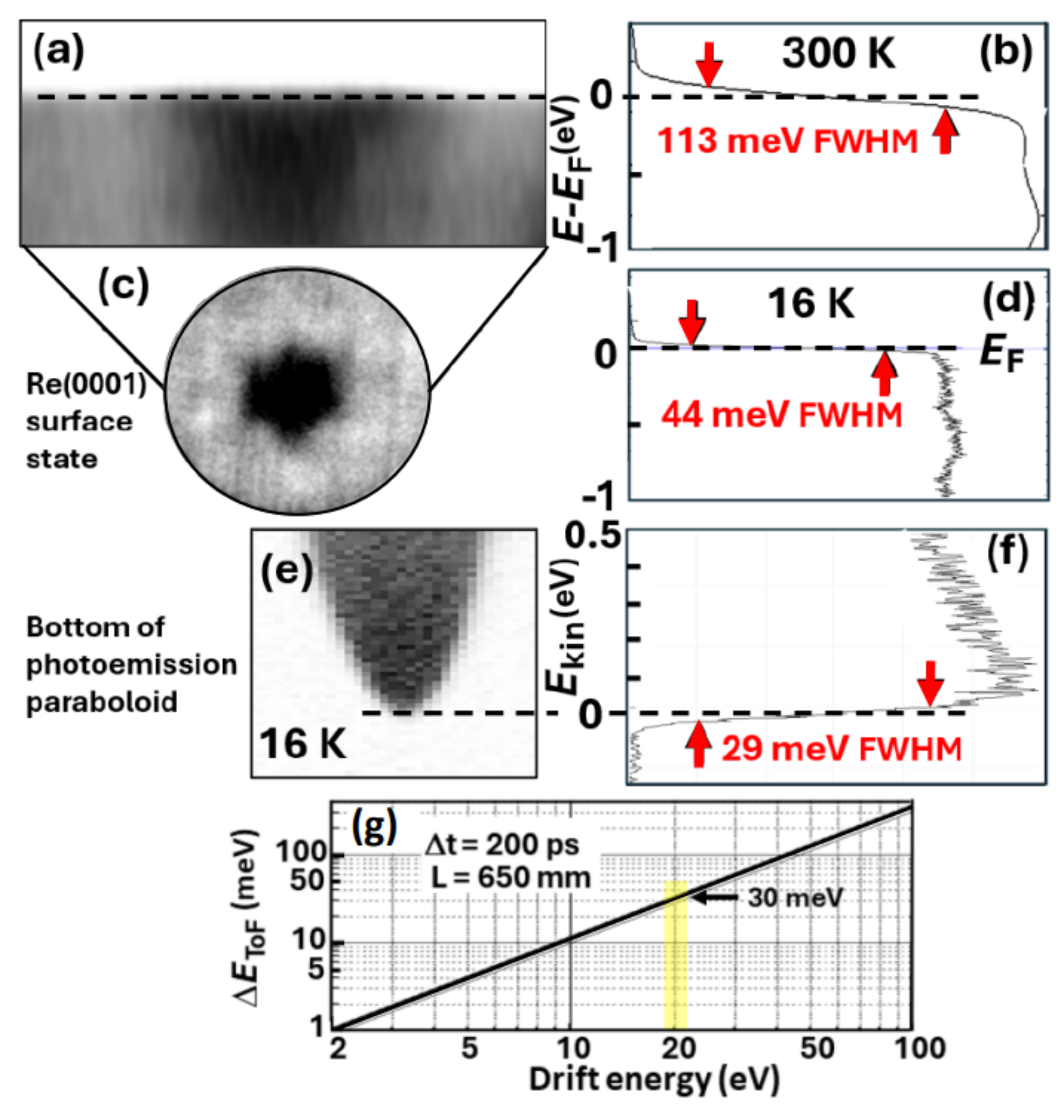}
\caption{\textbf{Determination of the energy resolution and XUV spectral bandwidth.} Measurement of the energy resolution at the Fermi edge and $E_{\text{kin}} = 0~\text{eV}$ cutoff of a Re(0001) sample; the drift energy in the ToF section was 20~eV. Momentum patterns ($E-E_F$)-vs-$k_x$ \textbf{(a)} and $k_x$-vs-$k_y$ \textbf{(c)} close to $E_F$ showing the Tamm surface state. \textbf{(b,d)} Intensity-vs-($E-E_F$) profiles for 300~K and 16~K, respectively. \textbf{(e)} Bottom of the photoemission paraboloid and \textbf{(f)} corresponding intensity profile showing the $E_{\text{kin}} = 0~\text{eV}$ cutoff. \textbf{(g)} Theoretical energy resolution of the ToF analyzer as a function of drift energy. The expectation of 30~meV at 20~eV agrees with the measurement \textbf{(f)}.}
\label{fig:energy}
\end{center}
\end{figure}

We have measured the total resolution $\Delta E_{\text{tot}}$ using photoemission from a metallic Re(0001) single crystal. The results are shown in Fig.~\ref{fig:energy} for the sample at room temperature (a)-(c) and at $16~\text{K}$ (d). The intensity profiles show cut-off widths of $113~\text{meV}$ and $44~\text{meV}$ FWHM, respectively. The Re(0001) single crystal had just been flashed before the measurement. The surface was not perfectly clean, so the band features of the Tamm surface state~\cite{Elmers20} appear diffuse in the momentum patterns, Figs.~\ref{fig:energy} (a),(c). However, the Fermi edge is not affected by such imperfections. The ToF energy resolution ($\Delta E_{\text{ToF}}$) can be conveniently measured at the bottom of the photoemission paraboloid, Figs.~\ref{fig:energy} (e),(f), since this low-energy cutoff is not thermally broadened. Figure~\ref{fig:energy} (f) shows the measured profile of this low-energy ($E_{\text{kin}} = 0$) cutoff, which has a width of $\Delta E_{\text{ToF}} = 29~\text{meV}$ FWHM. We can compare this value with the theoretical expectation for the drift energy of $E_{\text{d}} = 20~\text{eV}$ used for these measurements, given by the following equation~\cite{SPIECKER98}:

\begin{align}
	\label{eq:tof}
	\Delta E_{\text{ToF}} = \frac{2E_d}{m_e} \frac{L^2}{d\tau^2}
\end{align}

where $E_d$ is the drift energy, $m_e$ is the electron mass,  $L = 650~\text{mm}$ is the drift section length, and $d \tau = 200~\text{ps}$ is the delay line detector time resolution. Using these values, the calculated theoretical value is $30~\text{meV}$, in agreement with the measurement, see Fig.~\ref{fig:energy} (g). Inserting the values ($\Delta E_{\text{therm}} = 5.3~\text{meV}$ at $16~\text{K}$) yields a photon bandwidth of $\Delta E_{h\nu} = 33~\text{meV}$. Using this measured XUV spectral bandwidth allows us to calculate the Fourier-limit $\Delta t_{FL}$ of the XUV pulse duration using $\Delta E \Delta t_{FL} = 1825~\text{meV} \cdot \text{fs}$, which yields $\Delta t_{FL}=55~\text{fs}$. The XUV pulse duration is, however, expected to be typically a few 10$\%$ longer than this limit due to the inherent atomic dipole phase variation and associated frequency chirp of high-order harmonics emission~\cite{Mauritsson04, Mairesse05, Varju05}. 

\begin{figure}[t]
\begin{center}
\includegraphics[width=8.6cm]{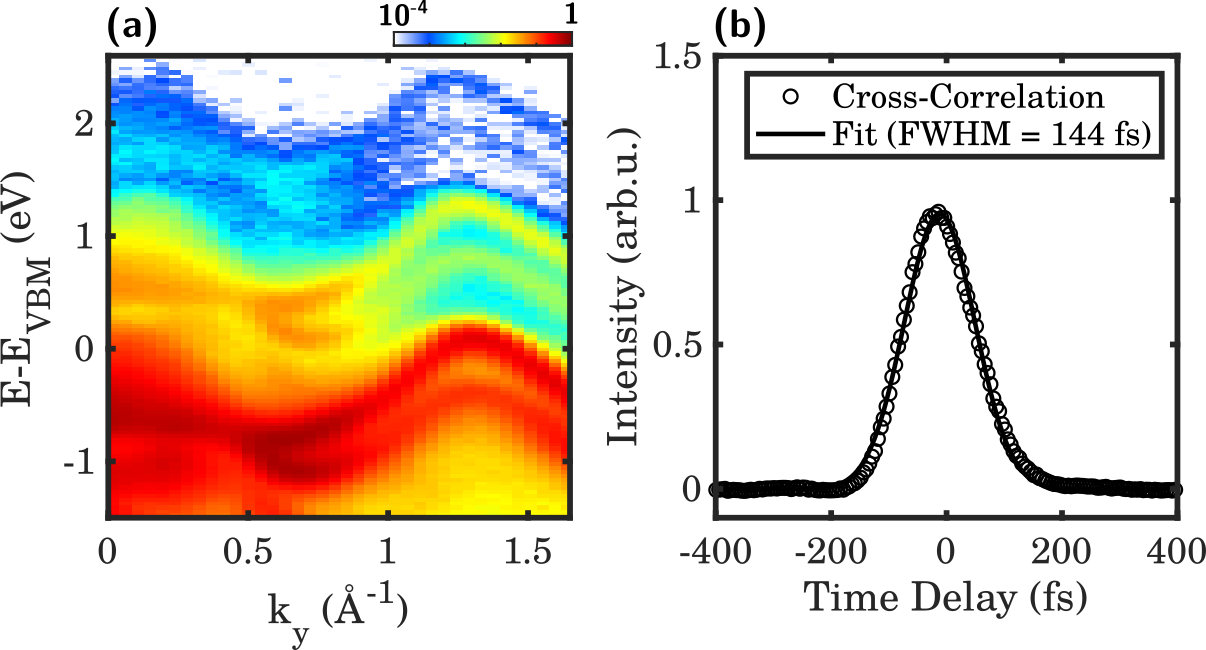}
\caption{\textbf{Characterization of the temporal resolution.} Using p-polarized driving IR pump pulse (1.2~eV) and p-polarized XUV probe, we measured the temporal evolution of Laser-Assisted PhotoEmission (LAPE) in 2H-WSe$_2$, which is a non-resonant coherent two-photon (XUV+IR) process allowing to experimentally determine the cross-correlation between the IR pump and XUV probe pulses. \textbf{(a)} Energy-momentum cut along $\Gamma$-K' high symmetry direction, integrated over all pump-probe delays (-400~fs to 400~fs). \textbf{(b)} Time-resolved LAPE signal and associated Gaussian fitting. From this measurement, the experimental temporal resolution is determined to be 144~fs FWHM.}
\label{fig:tempres}
\end{center}
\end{figure}

To characterize the temporal resolution of our apparatus, we measured the time-resolved Laser-Assisted PhotoEmission (LAPE) signal in 2H-WSe$_2$, which is a non-resonant coherent two-photon (XUV+IR) process. To do so, we used p-polarized driving IR pump pulse (1.2~eV), since this IR polarization state maximizes LAPE intensity~\cite{Mahmood2016-jv}. The time-resolved LAPE signal and associated Gaussian fitting are shown in Fig.~\ref{fig:tempres} (b). From this fitting procedure, we extracted an experimental temporal resolution of 144~fs FWHM. The temporal resolution of our experimental setup is limited by the duration of the IR pump pulse since the XUV pulse duration is expected to be shorter than the IR driver due to the extreme nonlinearity of the HHG process~\cite{Glover96, Varju05}.  

\section{\label{sec:dichro}Multi-Modal Photoemission Dichroism}

\begin{figure*}[t]
\begin{center}
\includegraphics[width=14cm]{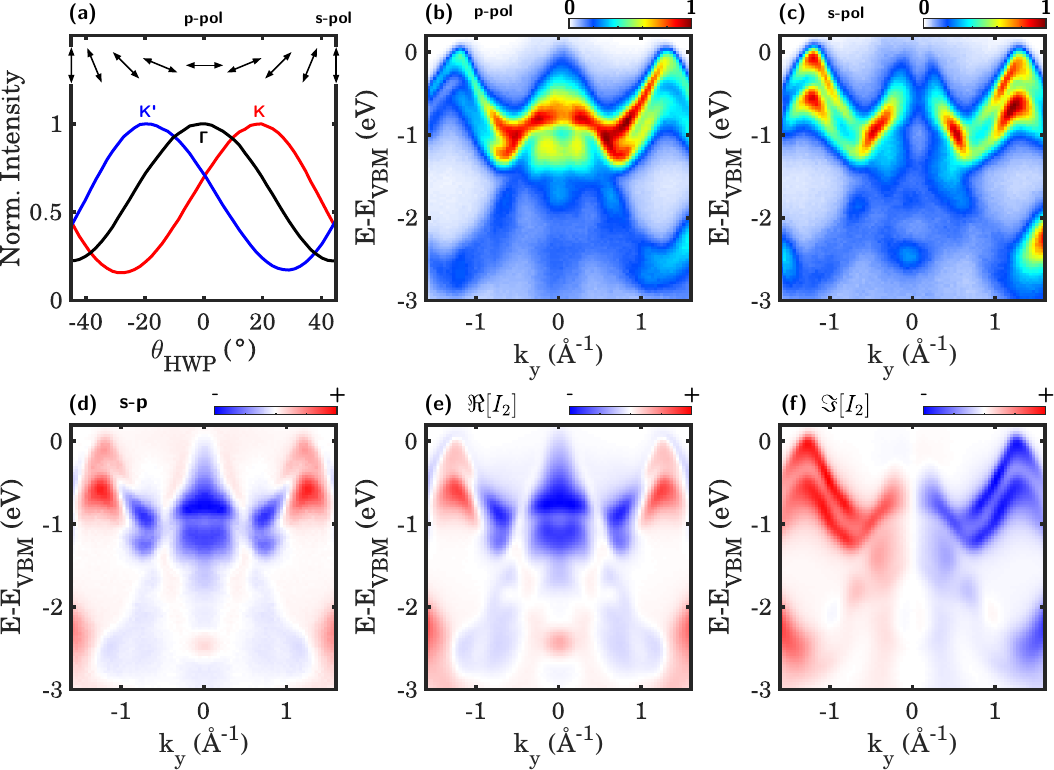}
\caption{\textbf{Extraction of linear and Fourier dichroism in 2H-WSe$_2$ from continuous XUV linear polarization axis angle modulation.} The light incidence plane is aligned with the crystal mirror plane, i.e. along the $\Gamma$-M direction. \textbf{(a)} Photoemission intensities as a function of the XUV half-wave plate angle ($\theta_\mathrm{HWP}$), around $\Gamma$ (black line), K valley (red line), and K$^{\prime}$ valley (blue line). \textbf{(b)-(c)} Energy-momentum cut along K-$\Gamma$-K' high symmetry direction for p-polarized and s-polarized XUV, respectively. \textbf{(d)} Linear dichroism in the photoelectron angular distributions (LDAD) calculated by subtracting the normalized photoemission signal using s- and p-polarized XUV. \textbf{(e)} Real part of the complex Fourier coefficient ($\Re[I_2]$) associated with the modulation of the photoemission intensity upon s- to p- polarization switch frequency, measured using continuous modulation of the XUV linear polarization axis angle. As described in the text, $\Re[I_2]$ is proportional to LDAD. \textbf{(f)} Imaginary part of the complex Fourier coefficient ($\Im[I_2]$) associated with the modulation of the photoemission intensity upon s- to p- polarization switch frequency measured using continuous modulation of the XUV linear polarization axis angle.}
\label{fig:ldad}
\end{center}
\end{figure*}

Now that we have presented a detailed characterization of the spatial, temporal, and energy resolutions of our instrument, we turn our attention to the main novelty of the apparatus: the multi-modal photoemission dichroism from light-induced nonequilibrium states in quantum materials.  In this section, we describe measurement methodologies behind the extraction of various types of dichroism using the polarization tunability of our XUV source in combination with the single-event electron detection capabilities of the momentum microscope. 

\subsection{Linear and Fourier dichroisms}
We start by introducing the measurement protocol to extract the linear and Fourier dichroisms using the continuous XUV linear polarization axis angle modulation, here exemplified for 2H-WSe$_2$. For these measurements, the all-reflective XUV quarter-wave plate is removed from the XUV beam path, and the light incidence plane is aligned with the crystal mirror plane, i.e. along the $\Gamma$-M direction. We continuously rotate the visible (515 nm) half-wave plate located before the HHG chamber while measuring the associated photoemission intensity on the fly. Each electron event is tagged with the instantaneous position of the half-wave plate. The polarization direction of the high-order harmonics is parallel to that of the driving laser. Therefore, when rotating the half-waveplate in the driving beam (515 nm), the linear polarization axis of the XUV light also rotates. We go through four cycles of p-, s-, p-polarization states, going through all intermediate linear polarization axis angles for the XUV radiation upon a full rotation (360$^{\circ}$) of the half-wave plate. We then bin the raw single-event data to generate a four-dimensional dataset, i.e. $I(E_\mathrm{B}, k_x, k_y, \theta_{HWP})$. The data presented in Fig.~\ref{fig:ldad} were recorded during $\sim$24 hours at a count rate $\sim$3E5 electrons/s, while continuously rotating the half-wave plate at a speed of 10$^{\circ}$/s. 

In Fig.~\ref{fig:ldad} (a), we show photoemission intensities as a function of the half-wave plate angle ($\theta_\mathrm{HWP}$), around the valence band maximum at $\Gamma$ (black line), K valley (red line), and K$^{\prime}$ valley (blue line). The photoemission intensity at $\Gamma$ is maximized for p-polarized XUV and cosinusoidally decreases when going towards s-polarized XUV. The situation is different for photoemission intensities from K and K$^{\prime}$ valleys. Indeed, at K valley, the photoemission signal goes through a maximum (minimum) at positive (negative) wave-plate angles, which correspond to intermediate polarization axis angles between s- and p-polarization. This behavior of the polarization-dependent photoemission yield is opposite for the K$^{\prime}$ valley. Fig.~\ref{fig:ldad} (b)-(c) shows the associated normalized energy-momentum cut along K-$\Gamma$-K$^{\prime}$ high symmetry direction for p- and s-polarized XUV, respectively. The linear dichroism computed by taking the difference between these two spectra is shown in Fig.~\ref{fig:ldad} (d). The signal has been symmetrized with respect to the $k_y = 0$ plane to remove small artifacts coming from imperfect experiment geometry effects. In this linear dichroism map, one can visualize the different polarization-dependence for $\Gamma$ and K/K$^{\prime}$ valleys. Indeed, while the signal is strong near $\Gamma$ with p-polarized XUV, switching to s-polarization leads to a redistribution of the photoemission intensity towards K/K$^{\prime}$ valleys. This behavior is visible from the positive (blue) and negative (red) linear dichroism at $\Gamma$ and K/K$^{\prime}$ valleys, respectively. While information about orbital characters and their in-plane/out-of-plane components can be extracted from such a linear dichroism map, it is evident that key details about the full polarization dependence of the photoemission yield from different states are missing. For instance, while the photoemission intensities at the K/K$^{\prime}$ valleys are equivalent when using s- and p-polarization states, they exhibit strikingly different behaviors when the polarization axis lies at intermediate angles. This behavior remains entirely undetectable using standard LDAD measurements, i.e. linear dichroism is the same at K and K$^{\prime}$. 

To extract the full information encoded in the polarization-dependence of the photoemission yield from different states, we perform a Fourier analysis of the periodic polarization-modulated signals, as described in Ref.~\cite{Schuler22}. Indeed, we perform an energy- and momentum-resolved Fourier analysis (for each energy-momentum voxel) along the XUV polarization axis to extract the real and imaginary parts of the oscillating photoemission yield. The Fourier-transformed signal
\begin{align}
	\label{eq:fourier_int}
	I_m(E,k_x,k_y) = \int^{2\pi}_0\frac{\mathrm{d}\theta}{2\pi}e^{i m \theta} I(E,k_x,k_y,\theta)
\end{align}
is only non-vanishing for $m=0,\pm 2$ coefficients. While $m=0$ corresponds to the polarization-averaged intensity (the zero frequency background), $I_2(E,k_x,k_y)$ is a complex quantity encoding the full information about photoemission modulation upon rotating the polarization axis angle of the XUV pulses. For our experimental geometry, we can link the real and imaginary parts of $I_2(E,k_x,k_y)$ to photoemission matrix elements, as in Ref.~\cite{Schuler22}:
\begin{align}
	\label{eq:i2_real}
	\Re[I_2(E,\vec{k})] \propto |M_s(E,\vec{k})|^2 - |M_p(E,\vec{k})|^2   \\
	\label{eq:i2_imag}
	\Im[I_2(E,\vec{k})] \propto  \Re\left[(M_s(E,\vec{k}))^* M_p(E,\vec{k})\right] \ , 
\end{align}
In Eq.~\ref{eq:i2_real} and \ref{eq:i2_imag}, $M_s(E,\vec{k})$ and $M_p(E,\vec{k})$ are the transition dipole matrix elements with respect to s- or p- polarized light. The real part $\Re[I_2(E,\vec{k})]$ is proportional to LDAD, which originates from the modulation of the photoemission intensity upon going from s- to p- polarized light. On the other hand, the imaginary part $\Im[I_2(E,\vec{k})]$ captures interference between these photoemission channels. Indeed, this interferometric quantity is sensitive to the relative phase between $M_s(E,\vec{k})$ and $M_p(E,\vec{k})$ in a very similar fashion to CDAD, as we will discuss later. As shown in Ref.~\cite{Schuler22}, this dichroic quantity $\Im[I_2(E,\vec{k})]$ encodes information about the electronic wavefunction that goes beyond standard linear dichroism. 

In Fig.~\ref{fig:ldad} (e)-(f), we show the extracted $\Re[I_2(E,\vec{k})]$ and $\Im[I_2(E,\vec{k})]$ signals along K-$\Gamma$-K' high symmetry direction. From now on, as in the label and caption of Fig.~\ref{fig:ldad}, we will drop the $(E,\vec{k})$ dependence notation for describing these quantities. As expected from Eq.~\ref{eq:i2_real}, one can notice from Fig.~\ref{fig:ldad} (e) that $\Re[I_2]$ is almost indistinguishable from the LDAD calculated by subtracting the normalized photoemission signal using s- and p-polarized XUV, shown in Fig.~\ref{fig:ldad} (d). The equivalence between the LDAD extracted from simple subtraction of photoemission intensities measured with s- and p-polarized XUV and from the Fourier analysis of the full continuous XUV linear polarization axis angle modulation scan confirms the validity of the approach. In Fig.~\ref{fig:ldad} (f), we can see that $\Im[I_2]$ exhibits a very simple and characteristic pattern: a strong signal with the same sign for the spin-orbit split VB1 and VB2 at K valley, which exhibits a sign change when going to the opposite valley pseudospin, i.e. K$^{\prime}$. As discussed in Ref~\cite{Schuler22}, this is reminiscent of the orbital pseudospin and Berry curvature texture of valence bands at K/K$^{\prime}$ valleys. Detailed discussion about the microscopic origin of the different types of dichroism in angle-resolved photoemission spectroscopy is beyond the scope of this manuscript. However, from Fig.~\ref{fig:ldad}, we can conclude that using the linear polarization axis tunability of our setup allows for the extraction of high-quality LDAD and $\Im[I_2]$ (also called Fourier Dichroism in the Photoelectron Angular Distributions - FDAD).

\begin{figure*}[t]
\begin{center}
\includegraphics[width=14cm]{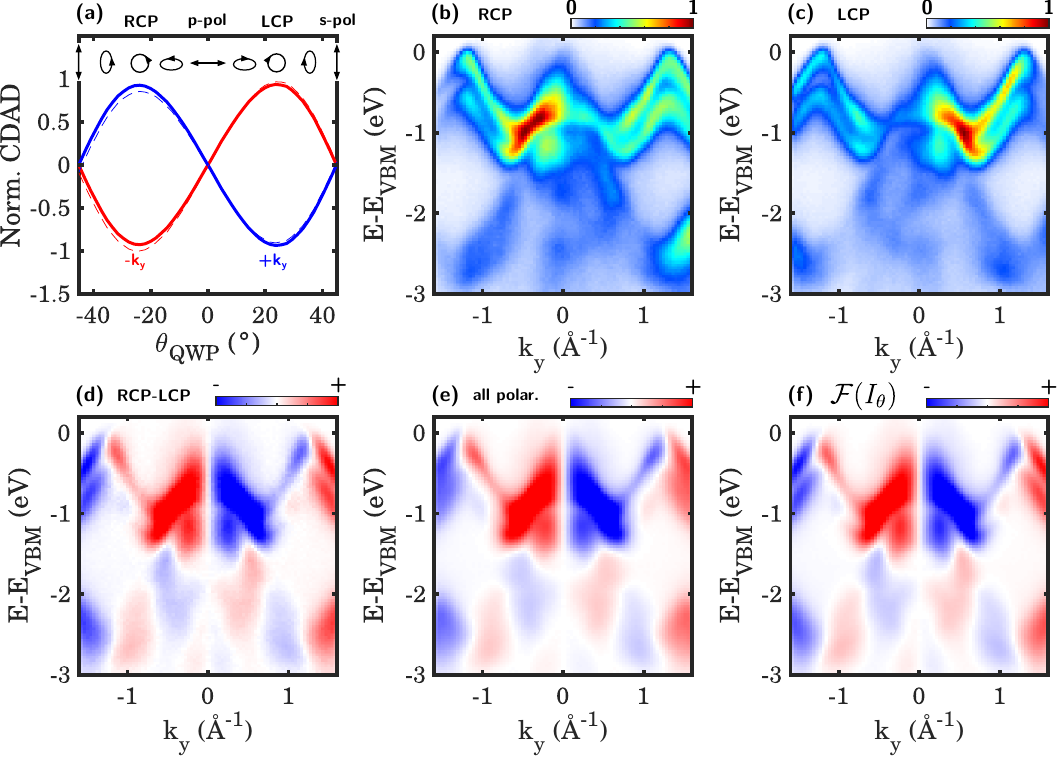}
\caption{\textbf{Extraction of circular dichroism (CD-ARPES or CDAD) in 2H-WSe$_2$ from continuous modulation of the polarization-state using XUV quarter-wave plate.} The light incidence plane is aligned with the crystal mirror plane, i.e. along the $\Gamma$-M direction. \textbf{(a)} Normalized CDAD as a function of linear polarization axis angle impinging on the XUV quarter-wave plate ($\theta_\mathrm{QWP}$), integrated for VB energy in positive ($+k_x$) and negative ($-k_x$) momenta along the K-$\Gamma$-K$^{\prime}$, in blue and red, respectively. CDAD is calculated by differentiating signals obtained from XUV polarization states with opposite helicities. Dashed lines represent raw CDAD while full lines are associated with CDAD antisymmetrized with respect to the $k_x=0$ plane. \textbf{(b)-(c)} Energy-momentum cut along K-$\Gamma$-K' high symmetry direction for quasi-RCP and quasi-LCP XUV polarization-state, respectively. \textbf{(d)} CDAD calculated by subtracting the normalized photoemission signal using quasi-RCP and quasi-LCP XUV polarization states. \textbf{(e)} CDAD averaged over all ellipticities. \textbf{(f)} CDAD obtained using Fourier analysis, extracting the modulation of the photoemission intensity at the helicity switch frequency.}
\label{fig:cdad}
\end{center}
\end{figure*}

\subsection{Circular dichroism}
We now turn our attention to the measurement protocol allowing to extract the circular dichroism (CDAD or CD-ARPES). For these measurements, we insert the all-reflective XUV quarter-wave plate in the XUV beam path by simply moving a translation stage onto which the four SiC mirrors are standing. The experimental geometry is unchanged, i.e. the light scattering plane is aligned with the crystal mirror plane, i.e. along the $\Gamma$-M direction of 2H-WSe$_2$. The methodology is extremely similar to the one previously described for the extraction of LDAD and FDAD. We continuously rotate the visible (515~nm) half-wave plate before the HHG chamber while measuring the associated photoemission intensity on the fly. Each electron event is tagged with the instantaneous position of the  half-wave plate. Since XUV pulses are now reflected on the quarter-wave plate, a full rotation (360$^{\circ}$) of the half-wave plate leads to four cycles of p-, quasi-LCP, s-, quasi-RCP, p-polarization state of the XUV radiation (with maximum ellipticities of $\sim$90$\%$~\cite{Comby22}). The raw single-event data are binned to generate a four-dimensional dataset, i.e. ${I(E_\mathrm{B}, k_x, k_y, \theta_{QWP})}$. The data presented in Fig.~\ref{fig:cdad} are also recorded for $\sim$24 hours at a count rate $\sim$3E5 electrons/s, while continuously rotating the half-wave plate at a speed of 10$^{\circ}$/s. 

In Fig.~\ref{fig:cdad} (a), we show the normalized CDAD as a function of the linear polarization axis angle impinging on the XUV quarter-wave plate ($\theta_\mathrm{QWP}$), integrated for VB energy in positive ($+k_y$) and negative ($-k_y$) momenta along the K-$\Gamma$-K$^{\prime}$ direction, in blue and red, respectively. The absolute value of CDAD is maximized for quasi-circular polarization states, goes through zero when using s- or p-polarized light and reaches intermediate values for intermediate elliptical polarizations. The CDAD changes sign for opposite helicities and when going from positive to negative (and vice versa) momenta. All these features are expected from simple symmetry considerations. Figure~\ref{fig:cdad} (b)-(c) shows the associated energy-momentum cuts along the K-$\Gamma$-K' high symmetry direction, for quasi-RCP and quasi-LCP XUV polarization-state, respectively. Using these data acquired by continuously rotating the linear polarization axis angle impinging on the XUV quarter-wave plate ($\theta_\mathrm{QWP}$), CDAD can be extracted in different fashions. The most intuitive and standard method is to simply subtract photoemission intensities measured using quasi-RCP and quasi-LCP XUV polarization states, as shown in Fig.~\ref{fig:cdad} (d). However, only a very small fraction of the full dataset is used for this CDAD extraction procedure, i.e. the data recorded for all intermediate ellipticies are not used. To overcome this drawback, we can compute the CDAD averaged over all ellipticities with opposite helicity, yielding the CDAD shown in Fig.~\ref{fig:cdad} (e). This signal is extracted from the full dataset and thus contains more statistics than the CDAD extracted only with quasi-RCP and quasi-LCP XUV polarization states. This enhanced statistic will become an important asset when extracting CDAD from weak light-induced excited states, for example. In addition, we can see that CDAD signals extracted using both procedures are remarkably similar. A more advanced and elegant way of extracting the CDAD from the full polarization-modulated photoemission intensities is through Fourier analysis. Indeed, similarly to the procedure previously described to extract $\Re[I_2]$ and $\Im[I_2]$, we can filter the polarization-modulated signal oscillating at the helicity-switch frequency, which also corresponds to CDAD. This type of Fourier measurement scheme is analogous to lock-in detection and allows isolating signals that are modulated at the helicity-switch frequency (CDAD), efficiently rejecting all other frequency components coming from linear dichroism, experimental geometry, or artifacts coming from imperfection of the polarization state. The CDAD extracted from Fourier analysis, labeled $\mathcal{F} (I_{\theta})$, is shown in Fig.~\ref{fig:cdad} (f), and is also remarkably similar to CDAD signals extracted using both previous procedures. 

Detailed discussion about subtle dichroic patterns and their microscopic origins is beyond the scope of this manuscript. However, it is important to discuss the general origin of CDAD with respect to transition dipole matrix elements and compare it to $\Re[I_2]$ and $\Im[I_2]$. Taking into account that CDAD is defined as the difference between photoemission intensity using RCP and LCP polarizations, and substituting these polarization vectors in the photoemission Fermi golden rule yields:

\begin{align}
	\label{eq:cdad}
	\mathrm{CDAD} (E,\vec{k}) \propto \Im\left[(M_s(E,\vec{k}))^* M_p(E,\vec{k})\right] \
\end{align}

where $M_s(E,\vec{k})$ and $M_p(E,\vec{k})$ are the transition dipole matrix elements with respect to s- or p- polarized light. From Eq.~\ref{eq:cdad}, one notices that CDAD captures interference between $M_s(E,\vec{k})$ and $M_p(E,\vec{k})$ photoemission channels, and is thus sensitive to their relative phase. As previously discussed in Ref.~\cite{Schuler22}, and by comparing Eq.~\ref{eq:cdad} and Eq.~\ref{eq:i2_imag}, we realize that CDAD and FDAD ($\Im[I_2]$) are complementary (real and imaginary) parts of the same complex quantity, i.e. $(M_s(E,\vec{k}))^{*} M_p(E,\vec{k})$. These dichroic signals are thus fundamentally linked, and this relationship can be exploited to reconstruct the pseudospin texture (complex band eigenvectors), in the case of two relevant bands, as in the vicinity of high-symmetry points in transition metal dichalcogenides (TMDCs), for example~\cite{Schuler22}.  

\begin{figure}[t]
\begin{center}
\includegraphics[width=8.6cm]{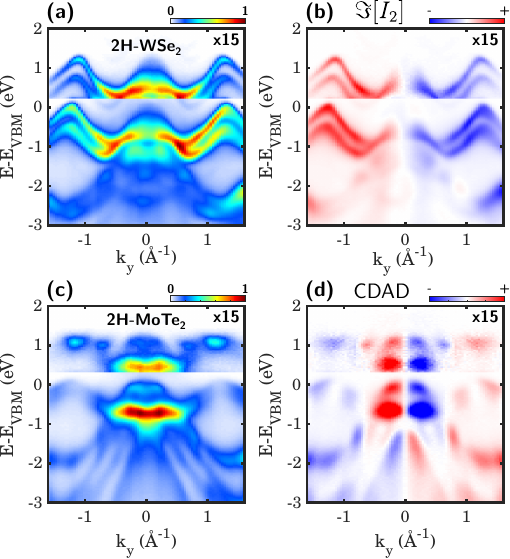}
\caption{\textbf{Fourier and circular dichroism from light-induced nonequilibrium states in 2H-WSe$_2$ and 2H-MoTe$_2$.} For both measurements, the light incidence plane is aligned with the crystal mirror plane, i.e. along the $\Gamma$-M direction. \textbf{(a)} Energy-momentum cut along K-$\Gamma$-K' high symmetry direction of 2H-WSe$_2$, using circularly-polarized IR pulses (135~fs, $\sim$5.7 mJ/cm$^2$), integrated over all XUV polarization from a continuous XUV linear polarization axis angle modulation scan. \textbf{(b)} Imaginary part of the complex Fourier coefficient ($\Im[I_2]$) associated with the modulation of the photoemission intensity upon s- to p- polarization switch frequency. \textbf{(c)} Energy-momentum cut along K-$\Gamma$-K' high symmetry direction of 2H-MoTe$_2$, using circularly-polarized IR pulses (135~fs, $\sim$2.6 mJ/cm$^2$), integrated over all XUV polarization from a continuous XUV quarter-wave plate scan. \textbf{(d)} CDAD obtained using Fourier analysis, extracting the modulation of the photoemission intensity at the helicity switch frequency.}
\label{fig:excited}
\end{center}
\end{figure}

\subsection{Dichroisms from nonequilibrium states}

Having established measurement methodologies to extract various types of dichroism using polarization-tunable XUV momentum microscopy, we go one step further and demonstrate the possibility of extending these measurements to light-induced nonequilibrium states. The general measurement protocol is the same as previously described, with the fundamental difference that we now spatially and temporally overlap an infrared (IR, 135~fs, 1030~nm) pump pulse with the polarization-tunable XUV probe pulse. We choose to present exemplary data from two different materials (2H-WSe$_2$ and 2H-MoTe$_2$), two flavors of dichroism (FDAD and CDAD), and two types of light-induced states (Volkov states, and conduction band population) to show the generality and the versatility of the approach. 

We start by investigating FDAD ($\Im[I_2]$) from Volkov states in 2H-WSe$_2$ (Fig.~\ref{fig:excited} (a)-(b)). To do so, we use a circularly-polarized IR pump pulse (135~fs, 1030~nm, $\sim$5.7 mJ/cm$^2$) temporally overlapped with the XUV probe pulse. We continuously scan the XUV linear polarization axis angle and measure the photoemission intensity on the fly, as previously described. The pump photon energy (1.2~eV) is much smaller than the bandgap of 2H-WSe$_2$ ($\sim$1.6~eV near K/K$^{\prime}$), preventing the population of excited states through single IR photon vertical transition. In Fig.~\ref{fig:excited} (a), we show an energy-momentum cut along K-$\Gamma$-K' high symmetry direction of 2H-WSe$_2$, obtained by integrating the photoemission signal over all XUV linear polarization axis angles. We can notice the emergence of Volkov states originating from the coherent interaction between the pump field and outgoing photoelectrons, and leading to a VB replica at $+\hbar\omega_\mathrm{IR}$. The associated FDAD signal is shown in Fig.~\ref{fig:excited} (b). The energy-momentum dependence of FDAD patterns from Volkov states is identical to VB' FDAD. This behavior is not surprising since the Volkov states originate from the interaction between the IR pump field and outgoing photoelectron and are thus expected to carry the same XUV polarization dependence as the VB. Even if the signal from Volkov is much weaker than the photoemission signal originating from the VB (x15 multiplication of the signal above VB maximum), FDAD from Volkov states has a very good signal-to-noise ratio, highlighting the capability of performing such dichroic measurements in nonequilibrium settings.   

We now turn our attention to CDAD from excited states in 2H-MoTe$_2$ (Fig.~\ref{fig:excited} (c)-(d)). To do so, we use circularly-polarized IR pump pulse (135~fs, 1030~nm, $\sim$2.6 mJ/cm$^2$) temporally overlapped with the XUV probe pulse. In this case, the pump photon energy (1.2~eV) is larger than the bandgap in 2H-MoTe$_2$ ($\sim$1.07~eV near K/K$^{\prime}$), leading to the population of excited states with IR photon vertical transitions. We continuously scan the XUV linear polarization axis angle in front of the all-reflective XUV quarter-wave plate and measure the photoemission intensity on the fly, as previously described. In Fig.~\ref{fig:excited} (c), we show an energy-momentum cut along K-$\Gamma$-K' high symmetry direction of 2H-MoTe$_2$, obtained by integrating the signal over all XUV polarization states. We can notice the presence of excited states signal in the K/K$^{\prime}$ and $\Sigma$ valleys, as well as relatively strong Volkov states near $\Gamma$. Weak Volkov signal also overlaps with excited state populations at the K/K$^{\prime}$ and $\Sigma$ valleys. The associated CDAD signal extracted using the Fourier analysis procedure previously described is shown in Fig.~\ref{fig:excited} (d). The CDAD from the valence band shows a sign reversal at the K and K$^{\prime}$ valleys. The VB CDAD near $\Gamma$ is strong and has an opposite sign with respect to K/K$^{\prime}$ valleys. The CDAD from Volkov states near $\Gamma$ is identical to the associated VB' CDAD. The dichroism from the excited states signals at K/K$^{\prime}$ has the same sign as the one from the VB. Again, a detailed discussion about subtle measured dichroic patterns and their microscopic origins is beyond the scope of this manuscript. Even if the signal from excited states is much weaker than the photoemission signal originating from the VB (x15 multiplication of the signal above VB maximum), CDAD from excited states has a very good signal-to-noise ratio. 

\subsection{Time- and valley-resolved XUV linear dichroism from photoexcited 2D materials}

\begin{figure}[t]
\begin{center}
\includegraphics[width=8.6cm]{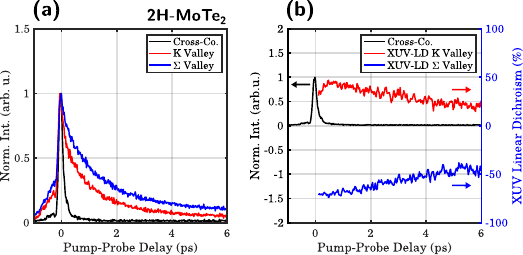}
\caption{\textbf{Time- and valley-resolved XUV linear dichroism from photoexcited 2H-MoTe$_2$.} \textbf{(a)} Normalized valley-resolved ultrafast population dynamics at K (red curve) and $\Sigma$ (blue curve) valleys in photoexcited 2H-MoTe$_2$ using circularly-polarized pump pulse ($\sim$2.6 mJ/cm$^2$), integrated for two pump-probe scans using s- and p-polarized XUV probe pulses. The black curve is the normalized cross-correlation signal between IR pump and XUV probe pulses. \textbf{(b)} Time-resolved XUV linear dichroism obtained by taking the normalized difference between photoemission signals measured using s- and p-polarized XUV probe pulses, at K (red curve) and $\Sigma$ (blue curve) valleys. The black curve is the normalized cross-correlation signal between IR pump and XUV probe pulses.} 
\label{fig:xuvld}
\end{center}
\end{figure}

Last, we demonstrate the feasibility of performing time-resolved XUV photoemission dichroism, exemplified here by XUV linear dichroism in photoexcited 2H-MoTe$_2$. To achieve this, we measured the energy- and momentum-resolved photoemission intensity as a function of pump-probe delay (ranging from -1 ps to 6 ps) for s- and p-polarized XUV probe pulses, with a circularly polarized pump pulse.

In Fig.~\ref{fig:xuvld}(a), we present the normalized valley-resolved ultrafast population dynamics at the K (red curve) and $\Sigma$ (blue curve) valleys following excitation by a 1.2 eV circularly polarized pump pulse at a fluence of $\sim$2.6 mJ/cm$^2$, integrated over two pump-probe scans using s- and p-polarized XUV probe pulses (photoemission intensity integrated between 0.8 and 1.2~eV above the valence band maximum). In transition metal dichalcogenides (TMDCs), valley-resolved ultrafast population dynamics at K and $\Sigma$ encode signatures of intra- and inter-valley scattering processes, which have been previously observed in time-resolved ARPES~\cite{Hein16, Bertoni16, dong21, madeo20, Kunin23}. These scattering processes are governed by the relative energy positions of different valleys, electron-phonon matrix elements, and electron-hole exchange interactions.

By taking the normalized difference between photoemission signals obtained using s- and p-polarized XUV probe pulses, we extract the time-resolved XUV linear dichroism in a valley-resolved manner [K (red curve) and $\Sigma$ (blue curve), Fig.~\ref{fig:xuvld}(b)]. To avoid any spurious artifacts from laser-assisted photoemission (LAPE), the time-resolved XUV linear dichroism is displayed only for positive pump-probe delays beyond the cross-correlation time range. A key observation is that the linear dichroism exhibits opposite signs at the K and $\Sigma$ valleys, reflecting the different orbital characters of the conduction bands at these high-symmetry points. Furthermore, while the time-resolved XUV linear dichroism at the $\Sigma$ valley decreases quasi-monotonically with increasing pump-probe delay, the dichroism at the K valley follows a more complex evolution: an initial increase on a subpicosecond timescale, followed by a slower decay. Although a detailed microscopic analysis of the ultrafast modulation of valley-resolved XUV linear dichroism is beyond the scope of this manuscript, our results demonstrate that such dichroic measurements encode dynamical information that cannot be accessed by standard analysis of population dynamics using a single probe polarization state.

\section{\label{sec:conclusion}Conclusions}

In conclusion, we reported the development and characterization of a novel time- and polarization-resolved extreme ultraviolet momentum microscopy apparatus. This instrument is specifically designed for time-resolved multimodal photoemission dichroism experiments on quantum materials driven out-of-equilibrium. The beamline is characterized by a time resolution of 144~fs and a total energy resolution of 44~meV (XUV spectral bandwidth of 33~meV FWHM). The XUV probe beam footprint on the sample is 45~$\mu$m x 33~$\mu$m (FWHM), which is relatively small compared to most time-resolved XUV ARPES setups. While the photon energy of the XUV probe arm is fixed at 21.6~eV, its polarization state is tunable (i.e. any linear polarization axis angle and quasi-circular). The momentum microscope endstation allows for simultaneous detection of the full surface Brillouin zone over an extended binding energy range and is equipped with improved electron optics with a new type of front lens that allows multiple modes of operation. Adjustable apertures can be inserted in the Gaussian and the reciprocal image planes allowing selection of small spatial regions-of-interest and enabling dark-field photoemission electron microscopy, respectively. The polarization tunability of our XUV source, coupled with the advanced detection capabilities of the momentum microscope, enables the study of linear, circular, and Fourier dichroism in photoelectron angular distributions from photoexcited samples. This paves the way for time-, energy-, and momentum-resolved investigation of wavefunction properties that govern the non-equilibrium behavior of quantum materials' electronic structures.

\section{Acknowledgement} We thank Nikita Fedorov, Romain Delos, Pierre Hericourt, Rodrigue Bouillaud, Laurent Merzeau, Sergey Babenkov, and Frank Blais for technical assistance. We acknowledge the financial support of the IdEx University of Bordeaux/Grand Research Program "GPR LIGHT". This work was supported by state funding from the ANR under the France 2030 program, with reference ANR-23-EXLU-0004, PEPR LUMA TORNADO. We acknowledge support from ERC Starting Grant ERC-2022-STG No.101076639, Quantum Matter Bordeaux, AAP CNRS Tremplin, and AAP SMR from Université de Bordeaux. SF acknowledges funding from the European Union’s Horizon Europe research and innovation programme under the Marie Skłodowska-Curie 2024 Postdoctoral Fellowship No 101198277 (TopQMat). Funded by the European Union. Views and opinions expressed are however those of the author(s) only and do not necessarily reflect those of the European Union. Neither the European Union nor the granting authority can be held responsible for them. 

\section{Availability of data} The data that support the findings of this article are openly available on Zenodo 
 \url{https://doi.org/10.5281/zenodo.17176237}.

%\bibliography{RSI_Setup}% Produces the bibliography via BibTeX.

%merlin.mbs aipnum4-1.bst 2010-07-25 4.21a (PWD, AO, DPC) hacked
%Control: key (0)
%Control: author (8) initials jnrlst
%Control: editor formatted (1) identically to author
%Control: production of article title (0) allowed
%Control: page (1) range
%Control: year (1) truncated
%Control: production of eprint (0) enabled
\providecommand{\noopsort}[1]{}\providecommand{\singleletter}[1]{#1}%

\end{document}